\documentclass[conference,a4paper]{IEEEtran}
\usepackage{graphicx}
\graphicspath{{./fig/}}
\usepackage{multirow}
\usepackage{rotating, graphicx}
\usepackage{array,multirow,graphicx}
\usepackage{amsmath}
\usepackage{amssymb}
\usepackage{comment}
\usepackage{lipsum}
\usepackage{mathtools}
\usepackage{float}
\usepackage{xcolor}
\usepackage[caption = false]{subfig}
\usepackage{amsmath} 
\usepackage[nolist]{acronym}
\usepackage{orcidlink}
\hypersetup{hidelinks}
\usepackage{colortbl,booktabs}
\usepackage{svg}

\makeatother
\usepackage{eso-pic}
\newcommand\AtPageUpperMyright[1]{\AtPageUpperLeft{
 \put(\LenToUnit{0.05\paperwidth},\LenToUnit{-1cm}){
     \parbox{1\textwidth}{\raggedleft\fontsize{9}{11}\selectfont #1}}
 }}
\newcommand{\conf}[1]{
\AddToShipoutPictureBG*{
\AtPageUpperMyright{#1}
}
}

\begin{document}

\title{Towards an AI/ML-driven SMO Framework in O-RAN: Scenarios, Solutions, and Challenges
\vspace{-2.9mm}}

\conf{This article has been accepted for publication in the IEEE Future Networks World Forum, Oct. 15-17, 2024, Dubai, United Arab Emirates.} 

\author{\IEEEauthorblockN{
Mohammad Asif Habibi\orcidlink{0000-0001-9874-0047}\IEEEauthorrefmark{1},
Bin Han\orcidlink{0000-0003-2086-2487}\IEEEauthorrefmark{1},
Merve Saimler\orcidlink{0000-0002-8447-7281}\IEEEauthorrefmark{3}, 
Ignacio Labrador Pavón\orcidlink{0000-0003-0652-5740}\IEEEauthorrefmark{4},
and Hans D. Schotten\orcidlink{0000-0001-5005-3635}\IEEEauthorrefmark{1}\IEEEauthorrefmark{2}}
\IEEEauthorblockA{\IEEEauthorrefmark{1}University of Kaiserslautern (RPTU), Germany; 
\IEEEauthorrefmark{3}Ericsson Research, Turkey;
\IEEEauthorrefmark{4}ATOS Research \& Innovation, Spain; \\ 
\IEEEauthorrefmark{2}German Research Center for Artificial Intelligence (DFKI), Germany
\vspace{-6.5mm}
 }} 

\maketitle

\begin{abstract}
The emergence of the \ac{O-RAN} architecture offers a paradigm shift in cellular network management and service orchestration, leveraging data-driven, intent-based, autonomous, and intelligent solutions. Within O-RAN, the \ac{SMO} framework plays a pivotal role in managing \acp{NF}, resource allocation, service provisioning, and others. However, the increasing complexity and scale of \acp{O-RAN} demand autonomous and intelligent models for optimizing \ac{SMO} operations. To achieve this goal, it is essential to integrate intelligence and automation into the operations of \ac{SMO}. In this manuscript, we propose three scenarios for integrating \ac{ML} algorithms into \ac{SMO}. We then focus on exploring one of the scenarios in which the \ac{Non-RT RIC} plays a major role in data collection, as well as model training, deployment, and refinement, by proposing a centralized \ac{ML} architecture. Finally, we identify potential challenges associated with implementing a centralized \ac{ML} solution within \ac{SMO}.
\vspace{0.7mm}
\end{abstract}

\IEEEoverridecommandlockouts
\begin{keywords}
3GPP, Artificial Intelligence, Data Analytics, ETSI, Management \& Orchestration, Machine Learning, O-RAN Architecture, O-RAN Alliance, Standards, SMO Framework
\vspace{-2.8mm}
\end{keywords}

\IEEEpeerreviewmaketitle

\begin{acronym}
\acro{AI}{artificial intelligence}
\acro{API}{Application Programming Interface}
\acro{CNF}{Containerized Network Function}
\acro{CSMF}{Communication Service Management Function}
\acro{E2E}{end-to-end}
\acro{SMO}{Service Management and Orchestration}
\acro{eMBB}{enhanced Mobile Broadband}
\acro{KPI}{Key Performance Indicators}
\acro{MANO}[M\&O]{management and orchestration}
\acro{ML}{Machine Learning}
\acro{MO}{Managed Object}
\acro{NF}{Network Function}
\acro{NFVO}{Network Function Virtualization Orchestrator}
\acro{NS}{Network Service}
\acro{QoS}{Quality of Service}
\acro{QoE}{Quality of Experience}
\acro{MDAS}{management data analytics service}
\acro{MDA}{management data analytics}
\acro{MDAF}{management data analytics function}
\acro{UPF}{User Plane Function}
\acro{WP6}{Sixth Workpackage}
\acro{O-RAN}{open radio access network}
\acro{SDO}{standards developing organization}
\acro{CNF}{containerized network function}
\acro{eMBB}{enhanced mobile broadband}
\acro{UE}{user equipment}
\acro{KPI}{key performance indicators}
\acro{ML}{machine learning}
\acro{MO}{managed object}
\acro{NF}{network function}
\acro{NFVO}{network function virtualization orchestrator}
\acro{NS}{network slice}
\acro{SLA}{service level agreement}
\acro{QoS}{quality of service}
\acro{SMO}{service management and orchestration}
\acro{Non-RT RIC}{non-real-time RAN intelligence controller}
\acro{Near-RT RIC}{near-real-time RAN intelligence controller}
\acro{NSS}{network slice subnet}
\acro{3GPP-NSMS}{3GPP-network slicing management system}
\acro{NFV-MANO}{NFV-management and orchestration}
\acro{O-RU}{open radio unit}
\acro{O-CU}{open centralized unit}
\acro{O-DU}{open distributed unit}
\acro{O-gNB}{open next generation node B}
\acro{O-Cloud}{open-cloud}
\acro{CPU}{central processing unit}
\acro{O-BH}{open-backhaul}
\acro{O-FH}{open-fronthaul}
\acro{O-MH}{open-midhaul}
\acro{RIC}{RAN intelligence controller}
\acro{3GPP}{Third Generation Partnership Project}
\acro{SDO}{standards development organization}
\acro{FCAPS}{fault, configuration, accounting, performance, security}
\acro{ETSI}{European Telecommunications Standards Institute}
\acro{ISG}{Industry Specification Groups}
\acro{TSG}{Technical Specification Group}
\acro{SA5}{Service and System Aspects Working Group 5}
\acro{NFV}{Network Function Virtualization}
\acro{NSMF}{network slice management function}
\acro{NSSMF}{network slice subnet management function}
\acro{SST}{slice/service type}
\acro{NST}{network slice template}
\acro{NSST}{network slice subnet template}
\acro{NFMF}{network function management function}
\acro{URLLC}{ultra-reliable and low-latency communications}
\acro{mMTC}{massive machine type communications}
\acro{V2X}{vehicle-to-everything}
\acro{HMTC}{high-performance machine type communications}
\acro{VNF}{virtual network function}
\acro{PNF}{physical network function}
\acro{FB}{functional block}
\acro{VIM}{virtualized infrastructure manager}
\acro{VNFM}{virtual network function manager}
\acro{WIM}{wide area network infrastructure manager}
\acro{CISM}{container infrastructure service management}
\acro{CIR}{container image registry}
\acro{CCM}{container infrastructure service cluster management}
\acro{CNF}{containerized network function}
\acro{VL}{virtual link}
\acro{CIS}{container infrastructure service}
\acro{NC}{network controller}
\acro{AAF}{Authentication and Authorization Function} 
\acro{SRDF}{Service Registration and Discovery Function}  
\acro{MF}{management function}
\acro{SMEF}{Service Management and Exposure Function}
\acro{DMEF}{Data Management and Exposure Function}
\acro{ENI}{Experiential Networked Intelligence}
\acro{CLA}{closed-loop automation}
\acro{GDPR}{General Data Protection Regulation}
\acro{SGD}{Stochastic Gradient Descent}
\acro{AIaaS}{AI as a service}
\acro{MLOps}{machine learning operations}
\acro{DataOps}{data operations}
\end{acronym}

\section{Introductory Remarks}\label{Sec:Introduction} 
\vspace{-1.5mm}
\IEEEPARstart{T}{he} rapid evolution of cellular networks towards open, virtualized, cloud-native, and slicing-aware architectures, exemplified by initiatives such as \ac{O-RAN}, has introduced a new era marked by flexibility, scalability, programmability, automation, and intelligence in network management and service orchestration \cite{10024837, alam2024comprehensive}. The pivotal driver of this paradigmatic transformation is \ac{SMO} within \ac{O-RAN}. \ac{SMO} encompasses \acp{MF} introduced by the \ac{O-RAN} Alliance, as well as those defined by other \acp{SDO} like the \ac{3GPP} and the \ac{ETSI} \cite{10056710}. These \acp{MF}, drawn from various \acp{SDO}, collectively provide \ac{MANO} services for \ac{O-RAN} components in a unified manner. \ac{SMO} acts as a central nervous system, intelligently managing \acp{NF}, autonomously orchestrating resources, and efficiently optimizing network performance within \ac{O-RAN} \cite{BuildingSMOFramework}. To that end, \ac{SMO} seamlessly integrates \ac{AI}/\ac{ML} algorithms into its operational architecture.

The integration of \ac{ML} algorithms, trained through big data analytics, into \ac{SMO} holds immense promise for enhancing network and energy efficiency, improving service quality and user satisfaction, enabling \ac{E2E} automation and full programmability, and facilitating proactive decision-making in dynamic and complex network environments \cite{10353004, 9112751}. By leveraging \ac{ML} algorithms, \ac{O-RAN} can unlock valuable insights from vast volumes of data, predict and mitigate potential issues, automate repetitive tasks to streamline operations, and make well-informed decisions \cite{10178010, 9627832}. The overarching objective of leveraging \ac{ML} is to empower predominantly autonomous and intelligent \ac{O-RAN} capable of self-configuration, self-monitoring, self-healing, and self-optimization with minimal human intervention \cite{9112751, 9627832}.

The considerable potential impact of \ac{AI}/\ac{ML} in \ac{O-RAN} has garnered significant research interest, as evidenced by various research papers. Some studies, such as \cite{10353004} and \cite{10103771}, propose general frameworks for the intelligentization and automation of \ac{O-RAN}, addressing various use cases enhanced by data analytics and intelligence. Other studies focus on specific aspects within \ac{O-RAN}, optimizing certain performance metrics. For example, research has delved into exploring scheduling policy for various types of \ac{O-RAN} slices \cite{9627832}, designing xApps for deep reinforcement learning-based closed-loop control of \ac{O-RAN} slicing \cite{9814869}, dynamic offloading of \ac{O-RAN} disaggregation to edge sites for local data processing and faster decision-making \cite{10352133}, proposing a programmable traffic control service model for \ac{O-RAN} \cite{10329918}, and addressing the placement and scaling of \ac{O-RAN} virtual \ac{NF} \cite{10175556}. Beyond applying \ac{AI}/\ac{ML} to the operations of \ac{O-RAN}, there have been efforts to explore specific \ac{ML} models for deploying and testing optimal intelligence and automation solutions within \ac{O-RAN}. Examples of such studies include those referenced in \cite{10077111} and \cite{9992174}. In \cite{10077111}, the authors propose an automated, distributed, and \ac{AI}-enabled testing framework designed to evaluate the decision-making performance, vulnerability, and security of \ac{AI} models deployed within \ac{O-RAN}. Finally, in \cite{9992174}, the authors introduce a framework that determines the data necessary for training an \ac{ML} model for \ac{O-RAN}, demonstrating significant performance improvement compared to traditional methods.

The analysis of the state-of-the-art reveals a predominant focus on devising general frameworks and assessing specific optimization solutions for integrating \ac{ML} into \ac{O-RAN}. However, we have recently studied the integration of \ac{MDA} into the management frameworks of \ac{3GPP} and \ac{ETSI}, as well as enhanced the \ac{Non-RT RIC} \ac{AI}/\ac{ML} capabilities, within the context of \ac{SMO} \cite{10604823}. This study was conducted based on a unified and standard-compliant \ac{SMO} framework proposed in \cite{BuildingSMOFramework}. To our knowledge, this was the first comprehensive study solely dedicated to the intelligentization and automation of \ac{SMO}.  Apart from our earlier work \cite{10604823}, it is crucial to note the limited focus thus far on harnessing \ac{ML} algorithms and intelligence to enhance the capabilities of \ac{SMO} within \ac{O-RAN}.

With the aim of delving deeper into the intelligentization and automation of \ac{SMO}, we build upon our previous work \cite{10604823} and make the following contributions in this article:

\begin{itemize}
   \item We \textbf{introduce three deployment scenarios for integrating \ac{ML} models into \ac{SMO}}. The primary objective of these scenarios is to allow \ac{SMO} to select an \ac{AI}/\ac{ML} integration solution (tailored to its deployment and behavior) from a range of \ac{AI}/\ac{ML} integration options. 
   \item We \textbf{concentrate on one of the proposed scenarios, wherein the \ac{ML} model undergoes training through a centralized \ac{ML} paradigm}. In this contribution, our objective is to introduce a comprehensive architectural solution, illustrating the \ac{E2E} workflow of an \ac{ML} model within the proposed centralized framework.
   \item We \textbf{identify several research challenges linked to the integration of a centralized \ac{ML} paradigm within the \ac{SMO} framework}. Addressing these challenges necessitates significant research endeavors to fulfill the escalating demands for enhancing the intelligence of \ac{SMO} and facilitating the automation and intelligentization of various use cases related to the \ac{MANO} of \ac{O-RAN} components.
\end{itemize}

\begin{figure*}[ht]
    \centering
    \includegraphics[height=4.5 in, width=7.15in]{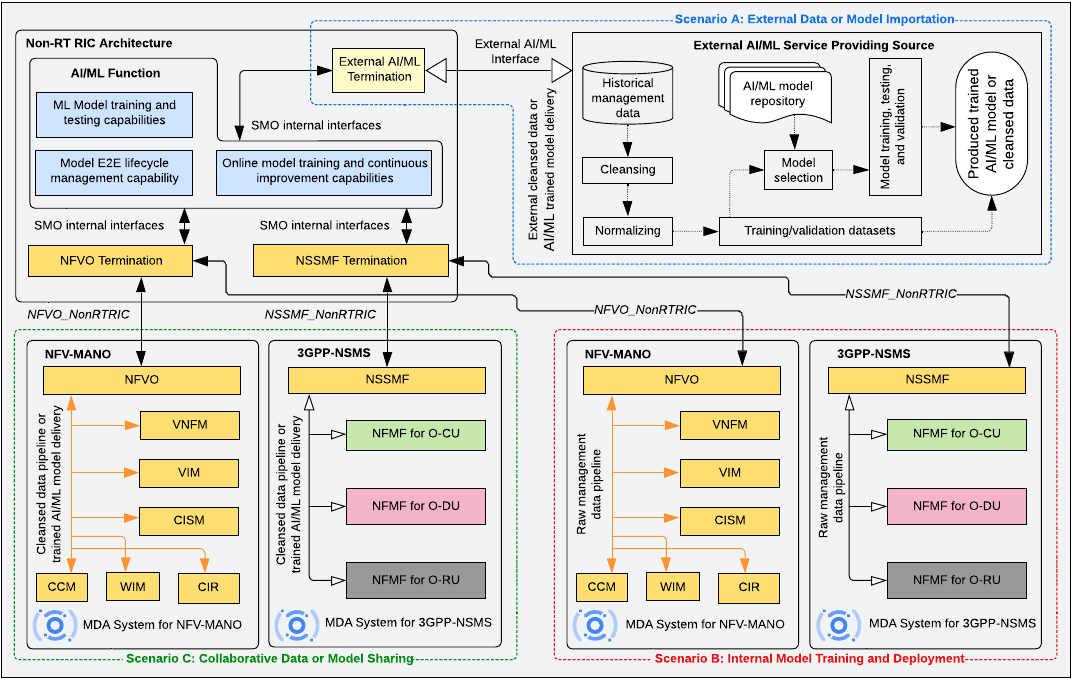}
    \caption{Three scenarios for incorporating AI/ML models into SMO are depicted. The scopes of Scenario A, Scenario B, and Scenario C are portrayed within the dashed blue, red, and green boxes, respectively. The legends contained within this figure have been defined in our previous work \cite{BuildingSMOFramework}. }
    \label{Fig:Scenarios}
    \vspace{-6mm}
\end{figure*}

The rest of this paper is organized as follows: We commence by introducing three scenarios for the training and deployment of \ac{ML} models within \ac{SMO} in Section \ref{Sec:Scenarios}. Then, we shift our attention to one of the scenarios, centralized \ac{ML}, highlighting the proposed architectural solution and procedures in Section \ref{Sec:InternalCentralized}. Next, we identify major research challenges associated with the centralized \ac{ML} solution in Section \ref{Sec:BenifitsandChallenges}. Finally, we summarize key points and identify promising directions for future research in Section \ref{Sec:Conclusions}.
\vspace{-0.5mm}

\section{Deployment Scenarios for Integrating MDA into the SMO Framework} \label{Sec:Scenarios}
\vspace{-1.5mm}
To integrate \ac{MDA}\footnote{In this paper, the terms ``\ac{MDA},'' ``intelligence,'' and ``\ac{AI}/\ac{ML} models'' are used interchangeably to denote the application of advanced computational techniques for extracting insights, making well-informed decisions, and optimizing management processes within \ac{O-RAN} in an intelligent and autonomous manner. Throughout the article, these terms convey the same meaning and are employed synonymously to emphasize the utilization of data-driven approaches for enhancing managerial intelligence within \ac{SMO}.} into \ac{SMO}, aiming to enhance its \ac{MANO} capabilities, we propose the consideration of a minimum of three scenarios for \ac{AI}/\ac{ML} model training, hosting, deployment, and monitoring, as shown in Figure \ref{Fig:Scenarios}. The three scenarios are outlined as follows: Scenario A: External Data or Model Importation; Scenario B: Internal Model Training and Deployment; and Scenario C: Collaborative Data or Model Sharing. These scenarios have been formulated upon the foundation of a unified \ac{SMO} we proposed in \cite{BuildingSMOFramework}, complemented by our research on the intelligentization and automation of \ac{SMO} in \cite{10604823}. In general, the three scenarios we present highlight various approaches to training, hosting, monitoring, and integrating \ac{AI}/\ac{ML} models into the operations of \ac{SMO}.

Each scenario offers its own set of advantages and challenges. Selecting the most suitable scenario hinges on various factors, including the specific use case, the expertise of network operators, the availability and confidentiality of data, as well as the volume of raw data to be exchanged between data-producing sources and model training components. Furthermore, determining an optimal \ac{AI}/\ac{ML} model training and deployment scenario for the automation and intelligentization of \ac{SMO} also depends on the metrics used to evaluate the performance of the respective \ac{AI}/\ac{ML} algorithm. 

The metrics used to evaluate the performance of an \ac{AI}/\ac{ML} model may include training time, inference time, cost and resource requirements, time-to-detection, time-to-resolution, and other relevant factors. Nevertheless, our unified and interoperable \ac{SMO} framework, presented in \cite{BuildingSMOFramework}, aims to provide a flexible and extensible architectural solution that supports the three previously mentioned deployment scenarios as well as other potential configurations that may be explored in the future. This capability enables seamless integration and interoperability with other standards-based \ac{AI}/\ac{ML} model providers, data-producing sources, and data-consuming components. In the remaining part of this section, we delve into each deployment scenario, providing a detailed discussion of its architecture, components, interfaces, and processes.

\subsection{Scenario A: External Data or Model Importation}
\vspace{-1.5mm}
In this scenario, the \ac{Non-RT RIC} imports cleansed data or a trained \ac{AI}/\ac{ML} model from an External \ac{AI}/\ac{ML} Service Providing Source via an External \ac{AI}/\ac{ML} Interface. The External \ac{AI}/\ac{ML} Services are provided by external components (i.e., beyond the scope of \ac{O-RAN}) and can be anchored outside of \ac{SMO} \cite{BuildingSMOFramework}. For example, the \ac{Non-RT RIC} framework imports a trained model or cleansed data from the \ac{ETSI} \ac{ENI} framework \cite{ETSIENIArchitecture} or \ac{AIaaS} providing platform, which can be situated outside of \ac{SMO}. The External \ac{AI}/\ac{ML} Service Provider must offer the necessary infrastructure and tools for data preprocessing, model training, and validation, aligning with best practices in \ac{MLOps}. It collects data from components within \ac{SMO}. Data collection can be achieved via standard-compliant \acp{API} to secure authorized access to the \ac{SMO} components and retrieve the necessary management data.

By identifying the relevant data sources and understanding the \ac{API} documentation, the External \ac{AI}/\ac{ML} Service Provider can make \ac{API} requests to gather the required management data and subsequently store it within an appropriate storage system. Following these steps, as depicted in Figure \ref{Fig:Scenarios}, the External \ac{AI}/\ac{ML} Service Provider can effectively preprocess and prepare management data for training. This involves crucial \ac{DataOps} practices such as data cleansing, normalization, and the creation of training and validation datasets. Upon completion of data preparation, the External \ac{AI}/\ac{ML} Service Provider proceeds with the selection and training of the \ac{AI}/\ac{ML} model, aligning it with the specific requirements of the use case. The model training process is conducted offline, utilizing the cleansed management data in a manner widely recognized as a best practice. The trained model and cleansed data must undergo rigorous testing and validation procedures before being deployed to the \ac{Non-RT RIC}. Upon completion of these processes, the cleansed data and the trained model are prepared for seamless deployment to the \ac{Non-RT RIC}.

The importation of either the externally cleansed data or the trained \ac{AI}/\ac{ML} model to the \ac{Non-RT RIC} is accomplished by utilizing the External \ac{AI}/\ac{ML} Termination via the External \ac{AI}/\ac{ML} Interface \cite{BuildingSMOFramework}, as illustrated in Figure \ref{Fig:Scenarios}. The External \ac{AI}/\ac{ML} Termination is a logical function that can be anchored within the architecture of the \ac{Non-RT RIC}. Upon successful importation of either the trained model or cleansed data, the \ac{AI}/\ac{ML} Function, which is a logical function and can also be anchored within the \ac{Non-RT RIC} framework \cite{10604823}, takes over sole responsibility for their \ac{E2E} lifecycle management. The decision of whether to import cleansed data or a trained model is a deployment choice made by the network operator.

In the event of importing cleansed data, the \ac{AI}/\ac{ML} Function assumes the responsibility for selecting, training, testing, and validating the appropriate \ac{AI}/\ac{ML} model before its execution. If the \ac{Non-RT RIC} imports the model, the \ac{AI}/\ac{ML} Function must validate the model to ensure it meets the requirements of the use case and adheres to all privacy, security, and regulatory standards. Once the model, whether imported or trained within the \ac{Non-RT RIC}, is prepared for execution, the \ac{AI}/\ac{ML} Function integrates it into the appropriate component(s) within \ac{SMO}, aiming to enhance the performance of an optimization function. In this scenario, the \ac{Non-RT RIC} may assume the responsibility for training (either online or offline), model selection, and lifecycle management, taking into consideration the imported management data. On the other hand, the \ac{Non-RT RIC}'s role in handling the imported \ac{AI}/\ac{ML} model may be limited to hosting, deployment, monitoring, and continuous improvement.

\subsection{Scenario B: Internal Model Training and Deployment}
\vspace{-1.5mm}
The \ac{Non-RT RIC} collects raw management data from various data-producing management systems and components within \ac{SMO} via standard-compliant interfaces. The \ac{Non-RT RIC} then cleanses and normalizes the collected data, as well as trains, tests, validates, executes, and monitors the selected \ac{AI}/\ac{ML} model. This specifically means that the \ac{AI}/\ac{ML} Function within the \ac{Non-RT RIC} collects and utilizes management data originating from the intelligent systems of the \ac{3GPP-NSMS}, \ac{NFV-MANO}, and other \ac{MANO} systems and components, as depicted in Figure \ref{Fig:Scenarios}. In this article, we adopt the \ac{MDA} System as the intelligent system of the \ac{3GPP-NSMS} and \ac{NFV-MANO}. Specific details regarding this adoption and the \ac{AI}/\ac{ML} Function are elaborated in \cite{10604823}.

The \ac{3GPP-NSMS} can encompass entities such as the \ac{NSSMF} and the \acp{NFMF} \cite{9750106}. The \ac{NFV-MANO} may incorporate several \acp{FB} and \acp{MF} \cite{9750106}, including the \ac{NFVO}, \ac{VNFM}, \ac{VIM}, \ac{WIM}, \ac{CISM}, \ac{CIR}, and \ac{CCM}. The \ac{Non-RT RIC} may also collect data from other \textit{de facto} and \textit{de jure} management systems and components that operate within \ac{SMO} and generate \ac{MANO} services for \ac{O-RAN}. The components of each intelligent system are interconnected with the components of its respective management system via standardized and interoperable interfaces \cite{10604823}.

The \ac{AI}/\ac{ML} Function of the \ac{Non-RT RIC} can collect the corresponding management data from the intelligent systems of the \ac{3GPP-NSMS} and \ac{NFV-MANO} through the use of \ac{NSSMF} Termination and \ac{NFVO} Termination via the \verb|NSSMF_NonRTRIC| and \verb|NFVO_NonRTRIC| interfaces, respectively. We initially introduced these two terminations and their relevant interfaces in our previous work \cite{BuildingSMOFramework}. We further expanded upon these terminations and their respective interfaces in another paper of ours referenced in \cite{10604823}, enhancing their capabilities to facilitate the exchange of training data and \ac{AI}/\ac{ML} models between the \ac{3GPP-NSMS}, \ac{NFV-MANO}, and \ac{Non-RT RIC} within \ac{SMO}. Once the data has been collected by the two terminations, the \ac{NSSMF} Termination and the \ac{NFVO} Termination seamlessly facilitate the transfer of collected data to the \ac{AI}/\ac{ML} Function via the \ac{SMO} internal interfaces.

Immediately upon receiving the relevant management data, the \ac{AI}/\ac{ML} Function commences the process of cleansing, normalizing, and preparing the data for validation, training, and testing. Subsequently, the \ac{AI}/\ac{ML} Function undertakes the training (offline or online) of the selected \ac{AI}/\ac{ML} model, hosts it, executes it within \ac{SMO}, continuously monitors its performance, and manages it throughout its operational lifespan. If the chosen model has successfully completed training and testing, the \ac{AI}/\ac{ML} Function expeditiously transfers the executable \ac{AI}/\ac{ML} model to the \ac{NSSMF} Termination and \ac{NFVO} Termination via designated \ac{SMO} internal interfaces.

The two terminations then take on the responsibility of conveying recommendations or predictions generated by the recently-trained \ac{AI}/\ac{ML} model through their respective interfaces to the management systems and components, namely the \ac{3GPP-NSMS} and \ac{NFV-MANO}. These management systems utilize the outputs of the trained model to improve the performance of a certain optimization function within \ac{SMO}. They also provide periodic reports to the \ac{AI}/\ac{ML} Function by supplying up-to-date management data. This process contributes to the improvement of the deployed \ac{AI}/\ac{ML} model through continuous online training. In this scenario, the \ac{Non-RT RIC} holds exclusive responsibility for the intelligentization and automation of the operations and maintenance of \ac{SMO}.

\subsection{Scenario C: Collaborative Data or Model Sharing} 
\vspace{-1.5mm}
In this scenario, the management data-producing components and systems within \ac{SMO} deliver either cleansed data or trained \ac{AI}/\ac{ML} models to the \ac{Non-RT RIC} via \ac{SMO} internal interfaces. The \ac{Non-RT RIC} then assumes the crucial responsibility of constructing a global model that enhances the performance of an optimization function within \ac{O-RAN}. It specifically means that the \ac{AI}/\ac{ML} Function either collects cleansed data or trained local \ac{AI}/\ac{ML} models from the \ac{MDA} System of the \ac{3GPP-NSMS} and the \ac{MDA} System of the \ac{NFV-MANO}, as shown in Figure \ref{Fig:Scenarios}. The \ac{AI}/\ac{ML} Function aggregates these individual local models or trains the cleansed data with the goal of constructing a global model.

The local models, also referred to as domain-level models in this article, are trained within specific management domains (i.e., \ac{NFV-MANO}, \ac{3GPP-NSMS}, and other systems within \ac{SMO}), utilizing the management data or network data stored on such domains. These models guarantee that confidential information remains stored on the respective management domains and is only transmitted when absolutely required or at specified time intervals, hence protecting user privacy and data security. The global model is trained using the cleansed data or trained local models stored within the central repository of data and models of the \ac{AI}/\ac{ML} Function. This model incorporates the collective learnings from cleansed data or all the local models of the \ac{MDA} Systems of the \ac{3GPP-NSMS} and \ac{NFV-MANO}. It acts as the coordinator, periodically aggregating updates from local models and incorporating them into its own parameters.

The \ac{MDA} System of the \ac{3GPP-NSMS} collects management data and network data from the \ac{NSSMF}, \acp{NFMF}, and the \acp{NF} of an \ac{O-gNB}. It undertakes data cleansing, conducts training for \ac{MF}-level/\ac{NF}-level and domain-level models, and subsequently executes these models to automate the operations of the corresponding \acp{MF} and the entire \ac{3GPP-NSMS}. \ac{MF}-level and \ac{NF}-level models are tailored models that are trained and executed specifically for a given \ac{MF} or \ac{NF}, respectively.
The trained models and cleansed data of the \ac{3GPP-NSMS} can be stored within its respective \ac{MDA} System for archival purposes. Depending on the configuration, the \ac{MDA} System may transfer either the cleansed data or the trained models to the \ac{NSSMF} Termination via the \verb|NSSMF_NonRTRIC| interface. The \ac{NSSMF} Termination may subsequently transfer the cleansed data or the trained models to the \ac{AI}/\ac{ML} Function via the \ac{SMO} internal interface. 

The \ac{MDA} System of the \ac{NFV-MANO} collects management data and network data from the \acp{MF}, \acp{FB}, and the underlying infrastructure. The \ac{MDA} System performs data cleansing, normalization, and data preparation tasks. Additionally, it engages in the training, validation, and testing processes for the corresponding models. Once the \ac{MF}-level, \ac{NF}-level, and domain-level models have been trained, the \ac{MDA} System executes them to intelligentize the operations of the \acp{FB}, \acp{MF}, \ac{NF}, and the entire \ac{NFV-MANO}. Based on the chosen design, the \ac{MDA} System of the \ac{NFV-MANO} may transfer either the cleansed data or the trained domain-level models to the \ac{NFVO} Termination via the \verb|NFVO_NonRTRIC| interface. The \ac{NFVO} Termination then forwards them via \ac{Non-RT RIC} internal interfaces to the \ac{AI}/\ac{ML} Function for additional processing.

If the \ac{Non-RT RIC} is configured to import cleansed data from the \ac{3GPP-NSMS} and \ac{NFV-MANO}, the \ac{AI}/\ac{ML} Function is then responsible for the validation, training, and testing of the global \ac{AI}/\ac{ML} model. Following the testing phase, it is anticipated that the \ac{AI}/\ac{ML} Function will develop an accurate and reliable global model. In the instance where the \ac{Non-RT RIC} imports trained domain-level models, the \ac{AI}/\ac{ML} Function constructs the global model by aggregating the \ac{3GPP-NSMS} and \ac{NFV-MANO} models. Once the global model (whether derived from imported cleansed data or domain-level trained models) is constructed and ready for deployment, the \ac{AI}/\ac{ML} Function provides it to the \ac{NFVO} Termination and \ac{NSSMF} Termination via the \ac{SMO} internal interfaces.

The two terminations provide the global model to the \ac{MDA} Systems of the \ac{NFV-MANO} and \ac{3GPP-NSMS} via the \verb|NFVO_NonRTRIC| and \verb|NSSMF_NonRTRIC| interfaces, respectively. The two \ac{MDA} Systems utilize the global model to enhance the performance and generalizability of their respective domain-level models. By providing access to a larger dataset, enforcing consistency, handling heterogeneity, correcting errors, and forming an ensemble, the global model can help domain-level models learn more accurately and generalize better to unseen data. This collaborative approach between domain-level and global models is a key enabler of federated learning, allowing us to develop \ac{AI}/\ac{ML} models that are both privacy-preserving and highly effective.

In Scenario C, the \ac{Non-RT RIC} holds a partial role in the intelligentization and automation of the operations associated with \ac{SMO}. The scope of the \ac{Non-RT RIC} might be restricted to that of Scenario A. Hence, continuous interactions among the components and interfaces of \ac{SMO} are essential to facilitate enhanced \ac{AI}/\ac{ML} model delivery and refinement.

\section{Centralized ML Model Training and Deployment Solutions within SMO} \label{Sec:InternalCentralized}
\vspace{-1.0mm}
Following the introduction of three scenarios for integrating \ac{ML} models into \ac{SMO}, we now focus on Scenario B. In this scenario, data is gathered at a single centralized location -- the \ac{Non-RT RIC}. The selected model is trained using this data, deployed across \ac{SMO}, and continuously refined. In this section, we outline a detailed workflow for realizing centralized \ac{ML} model training within the \ac{Non-RT RIC} and its deployment across \ac{SMO}. The proposed workflow consists of three phases, as shown in Figure \ref{Fig:CentralizedWorkflow}: (a) data collection and preprocessing; (b) model training and development; and (c) model deployment and inference. By conscientiously following the three phases, robust and effective \ac{ML} models can be developed that deliver valuable results for the intended optimization problem. In the following, we discuss these three phases, respectively.

\subsection{Data Collection and Preprocessing}\label{Subsec:DataCollection}
\vspace{-1.5mm}
This phase involves gathering and organizing the requisite data for training a selected \ac{ML} model within the \ac{Non-RT RIC}. The data is collected at the \ac{AI}/\ac{ML} Function from diverse sources located within the \ac{3GPP-NSMS}, \ac{NFV-MANO}, \ac{Non-RT RIC} itself, and potentially other systems reside within \ac{SMO} via standard-compliant interfaces. The \ac{Non-RT RIC} can employ techniques such as streaming and batch collection to capture both real-time and historical data from the aforementioned sources. The collected data primarily encompasses real-time operational data pertaining to network management, service orchestration, and \ac{SMO} performance.

Within the \ac{3GPP-NSMS}, the \acp{NFMF} hold a rich set of management data related to the \ac{MANO} of \ac{O-CU}, \ac{O-DU}, and \ac{O-RU}. These \acp{NFMF} collect and provide such data to the \ac{NSSMF}. The \ac{NSSMF} holds its own data and that of its associated \acp{NFMF}. We assume that the \ac{NSSMF}, alongside its fundamental functionalities, possesses the capability to generate and utilize \ac{AI}/\ac{ML} services within the \ac{3GPP-NSMS}. Based on this assumption, the \ac{NSSMF}, on behalf of the \ac{MDA} System of the \ac{3GPP-NSMS}, provides raw management data via the \verb|NSSMF_NonRTRIC| interface to \ac{NSSMF} Termination. Subsequently, this data is transferred to the \ac{AI}/\ac{ML} Function via \ac{SMO} internal interfaces for further processing.

The \acp{FB} and \acp{MF} within the \ac{NFV-MANO}, as discussed in the preceding section, contain management data associated with the virtual and cloud-native aspects of an \ac{O-gNB} and the underlying wireless infrastructure. Each \ac{MF} and \ac{FB} collects and holds its respective data. This data can then be transmitted to and stored within the \ac{NFVO}. Assuming that the \ac{NFVO}, in addition to its core functionalities, possesses the capability to produce and consume \ac{AI}/\ac{ML} services. Hence, under the authority of the \ac{MDA} System of the \ac{NFV-MANO}, the \ac{NFVO} furnishes management data pertaining to the \ac{NFV-MANO} via the \verb|NFVO_NonRTRIC| interface to the \ac{NFVO} Termination. Furthermore, this data is conveyed to the \ac{AI}/\ac{ML} Function via \ac{SMO} internal interfaces for subsequent treatment.

The \ac{AI}/\ac{ML} Function gathers relevant data from the components of the \ac{Non-RT RIC} framework and the rApps via the R1 interface and the internal interfaces of the \ac{Non-RT RIC}. In addition, the \ac{AI}/\ac{ML} Function can collect corresponding data from other de facto and de jure management systems within \ac{SMO} via internal interfaces specific to the \ac{SMO}.

\begin{figure}[ht]
    \centering
    \includegraphics[height=4.81 in, width=3.50in]{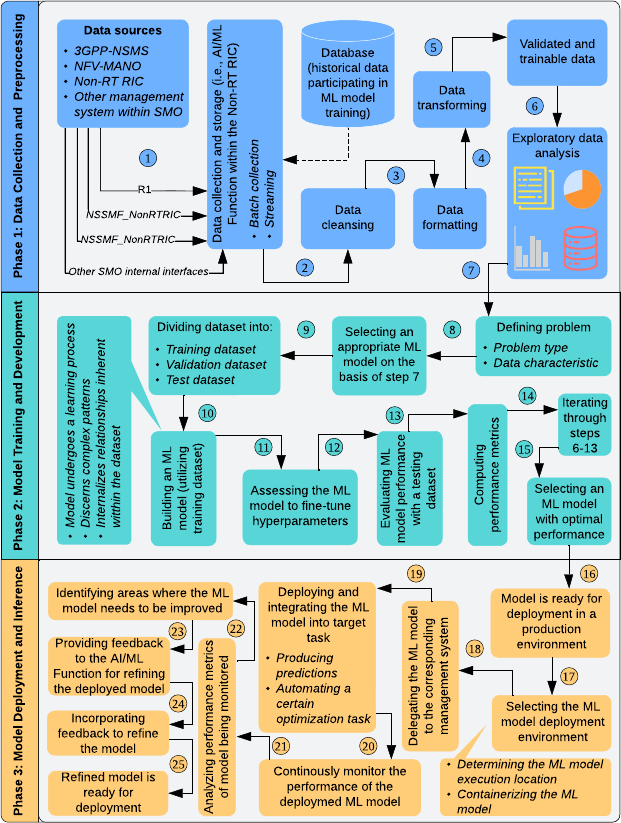}
    \caption{Proposed workflow for centralized ML model training, deployment, and refinement within the SMO framework of O-RAN architecture}
    \label{Fig:CentralizedWorkflow}
\end{figure}

Once the data has been completely gathered from the aforementioned data sources and stored within the \ac{AI}/\ac{ML} Function, it may contain errors, inconsistencies, and irrelevant information, necessitating preprocessing and polishing. Data preprocessing addresses these critical issues by cleansing, formatting, and transforming the collected data into a usable format for the \ac{ML} model, as shown in Figure \ref{Fig:CentralizedWorkflow}. During the cleansing process, the \ac{AI}/\ac{ML} Function removes duplicated data, corrects errors, and handles missing values. In the formatting process, it standardizes collected data into a common format, facilitating comparisons. In the transformation process, the \ac{AI}/\ac{ML} Function scales features, converts categorical data to numerical data (i.e., encoding), and performs feature engineering (i.e., creating new features from existing ones).

Following the above procedures, the \ac{AI}/\ac{ML} Function undertakes exploratory data analysis on preprocessed data, as shown in Figure \ref{Fig:CentralizedWorkflow}. During this process, various techniques such as visualization and statistical analysis are utilized to delve into the dataset, aiming to understand crucial variables, their correlations, magnitudes, and statistical attributes concerning the target variable. This process serves to pinpoint potential issues, guide decisions throughout model development, and distinguish between valuable features and irrelevant ones within the dataset. By executing these procedures, the data is assumed to be prepared for subsequent processes and consequently delivered to the second phase within the proposed workflow, as discussed in the following and illustrated in Figure \ref{Fig:CentralizedWorkflow}.

\subsection{Model Training and Development}\label{Subsec:ModelTraining}
\vspace{-1.5mm}
Once the collected data has undergone preprocessing and validation, and has been successfully received by this phase, the \ac{AI}/\ac{ML} Function proceeds to initiate the training and development of a selected \ac{ML} model. The Model Training and Development phase encompasses crucial processes in \ac{ML} model lifecycle management, including appropriate model selection, training, and development, as illustrated in Figure \ref{Fig:CentralizedWorkflow}. Each process within this phase holds significant importance in crafting resilient and efficient \ac{ML} models tailored to address specific optimization problems within \ac{O-RAN}.

During the model selection process, the initial crucial step is to clearly define the optimization problem that the model aims to address, as depicted in Figure \ref{Fig:CentralizedWorkflow}. A comprehensive understanding of the optimization problem thoroughly guides the selection of appropriate \ac{ML} algorithms. Depending on the problem type, data characteristics, and computational resources available, a set of candidate \ac{ML} models is chosen. Common choices encompass linear regression, support vector machines, decision trees, and neural networks, among many others. Due to privacy and security concerns, the \ac{AI}/\ac{ML} Function is also responsible for determining whether an \ac{AI}/\ac{ML} model shall be deployed in a federated or centralized manner before starting to collect data from data sources.

Selecting an optimal centralized \ac{ML} model requires a robust evaluation strategy. This entails dividing the data into training, validation, and test sets, as shown in Figure \ref{Fig:CentralizedWorkflow}. The model is then trained on the training set and assessed on the validation set to fine-tune hyperparameters, which are the configurations of the model. Finally, the performance of the model is evaluated on the unseen test set. To that end, a common set of evaluation metrics includes accuracy for classification tasks or mean squared error for regression tasks.

During the process of training an \ac{ML} model, the selected model is subjected to training using a training dataset. Throughout this training step, the model undergoes a learning process wherein it discerns and internalizes the complex patterns and relationships inherent within the data. Hyperparameter tuning emerges as a critical step in enhancing the performance of the selected model.

Within the context of \ac{ML}, hyperparameters are configurations that govern the learning process of an \ac{ML} model, such as the learning rate or the depth of a decision tree. Adjusting these hyperparameters can significantly impact the effectiveness and efficiency of the model. Techniques like grid search and randomized search are commonly employed to explore various combinations of hyperparameters, identifying the optimal settings that result in superior performance on the validation set during the training process.

In the model development process, the performance of the trained \ac{ML} model is carefully evaluated on the hold-out test set. This step provides an unbiased estimate of how well the model will generalize to unseen data. Based on the evaluation metrics, the best-performing \ac{ML} model from the candidate pool is selected. Furthermore, understanding how the model arrives at its predictions can be crucial. Techniques such as feature importance analysis help identify which features have the most significant influence on the model's output. This insight can provide a valuable understanding of the target optimization problem domain within \ac{O-RAN}, aiding in informed decision-making and \ac{AI}/\ac{ML} model refinement.

\subsection{Model Deployment and Inference}
\vspace{-1.5mm}
If the trained \ac{ML} model performs optimally, it can be deployed into a production environment for real-time or near-real-time inference. The production environments include the \ac{3GPP-NSMS}, \ac{NFV-MANO}, and other systems within the \ac{SMO}. The \ac{AI}/\ac{ML} Function performs the model deployment and inference via the \verb|NSSMF_NonRTRIC|, \verb|NFVO_NonRTRIC|, R1, and other standard-compliant interfaces. This phase entails integrating the \ac{ML} model into the aforementioned management systems, where it can be utilized for real-time predictions or task automation. Additionally, this phase may involve a refinement process wherein the performance of the deployed model is continuously monitored and periodically retrained with new data to ensure its continued effectiveness in addressing optimization tasks within \ac{O-RAN}.

In this phase, several critical steps must be followed, as illustrated in Figure \ref{Fig:CentralizedWorkflow}. First, the selection of a deployment environment is paramount. This entails deciding whether the trained model will reside on on-premises servers, cloud platforms, or edge devices. The location where the model is executed depends on the requirements of the optimization problem. For specific services like \ac{URLLC}, it may be necessary for the model to be executed at the edge, such as at the \ac{NFMF}, to effectively optimize the management of these services. Following determining the deployment environment, containerization becomes a crucial consideration, though optional. Packaging the \ac{ML} model, along with its dependencies, into a container like Docker ensures consistency across diverse environments, facilitating seamless deployment transitions. Scalability considerations form another pivotal aspect of deployment planning. Analyzing factors like anticipated workload, concurrent requests, and resource utilization aids in devising a deployment strategy that accommodates varying levels of demand \cite{9750106}.

Furthermore, robust monitoring and refining mechanisms are essential for optimizing the performance of the deployed model. Implementing a monitoring system within the \ac{AI}/\ac{ML} Function to track input data, predictions, and encountered errors provides valuable insights for continual improvement and refinement of the deployed model. Before refining the model, it is essential to assess its current performance using relevant evaluation metrics. These metrics could vary depending on the specific optimization problem as well as the characteristics and objectives of the model. Conducting a thorough analysis to identify areas where the model may be underperforming or exhibiting undesirable behavior is of vital importance. This analysis may involve examining misclassified samples, exploring feature importance, or analyzing model biases. 

Once the \ac{AI}/\ac{ML} Function receives feedback, the model can embark on a second attempt to enhance the outcome, as depicted in Figure \ref{Fig:CentralizedWorkflow}. In this phase, the algorithm incorporates feedback from the initial iteration and initiates improvements accordingly. As the model undergoes refinement, it yields a more polished output, thereby aiding in training the system to better comprehend real-world scenarios and expectations. Finally, by carefully navigating through these steps, a well-deployed model can fulfill its intended purpose effectively and securely, instilling confidence in its reliability and integrity.


\section{Major Research Challenges}\label{Sec:BenifitsandChallenges}
\vspace{-1.5mm}
This section outlines several research challenges linked to implementing \ac{AI}/\ac{ML} solutions centrally within \ac{SMO}. Addressing these challenges demands considerable research efforts to align with the escalating demands of future-oriented \ac{SMO} frameworks. It also aims to facilitate the \ac{MANO} of the emerging use cases and applications within \ac{O-RAN}.

\subsection{Privacy} 
\vspace{-1.5mm}
The first technical challenge arising from the \ac{SMO}-centralized \ac{AI}/\ac{ML} solution is data privacy. In Scenario B, the raw management data collected from various sources within \ac{SMO} may contain sensitive information, such as non-access stratum signaling data. Consistently transmitting such data to the \ac{Non-RT RIC} and maintaining it therewith for model training and development poses an increased risk of data exposure and privacy breaches, particularly in scenarios where midhaul connections are established wirelessly. To address this challenge effectively, the \ac{AI}/\ac{ML} Function must implement robust data privacy mechanisms to ensure that sensitive information is protected. These mechanisms may include data anonymization, encryption, and secure data transmission protocols. Furthermore, the \ac{AI}/\ac{ML} Function must adhere to data protection regulations, such as the \ac{GDPR} in the European Union, to safeguard user privacy throughout the data processing lifecycle.

\subsection{Resilience} 
\vspace{-1.5mm}
Ensuring the resilience of the centralized framework, particularly against data poisoning attacks, emerges as another critical concern. Centralized learning methodologies are inherently vulnerable to compromised raw data \cite{SMA+2024resilience}. In the context of \ac{SMO}, the raw data to train the \ac{AI}/\ac{ML} model may be tampered with, potentially leading to inaccurate predictions and suboptimal performance. Possible approaches to realizing such tampering include hijacking or manipulating user equipment to generate abnormal network behavior, or directly injecting malicious data packets into the raw management data pipeline or database. To effectively address this challenge, the \ac{AI}/\ac{ML} Function must implement robust data validation mechanisms to detect and filter out compromised data.

\subsection{Elasticity}
\vspace{-1.5mm}
Commonly in practice, the centralized \ac{AI}/\ac{ML} models must be occasionally retrained to keep up with the evolving network conditions and user requirements. However, such retraining often requires significant computational resources. By sharing the resources of a \ac{Non-RT RIC} with other \acp{NF}, the retraining procedures may trigger resource contention and degrade network performance. To achieve optimal elasticity, it calls for well-designed virtual resource scheduling at the \ac{Non-RT RIC} that can adapt to changing workload patterns and adaptively prioritize the needs of the \ac{AI}/\ac{ML} Function and other \acp{NF}.

\subsection{Agility}
\vspace{-1.5mm}
In addition to elasticity, the agility of the centralized \ac{AI}/\ac{ML} model emerges as another critical challenge, particularly concerning model retraining. While the model is trained using a vast amount of data aggregated from different domains, which usually takes a long time to converge, the network conditions, traffic patterns, and user requirements may change rapidly in individual local areas. It becomes therefore a crucial issue, how the centralized model in \ac{AI}/\ac{ML} Function can quickly adapt to the dynamic local environments, ensuring its continued effectiveness and accuracy in real-time scenarios. To tackle this challenge, advanced approaches are essential to enable the model to continuously and agilely learn and adapt to local variations and dynamics. This may involve incremental learning techniques, such as \ac{SGD} \cite{9461192}, which allow to efficiently update a trained model based on only the differential data, instead of retraining the model from sketch on with the entire dataset. Nevertheless, this may also introduce new challenges related to incremental learning techniques, e.g., model drift and catastrophic forgetting.

\subsection{Signaling Efficiency}
\vspace{-1.5mm}
Even with optimal learning efficiency that resolves the challenges of elasticity and agility, the inherent design of the centralized \ac{AI}/\ac{ML} solution, which involves aggregating massive amounts of raw data, gives rise to a significant signaling overhead. This overhead can potentially lead to network congestion at the midhaul, consequently causing heightened latency and diminished \ac{QoS}.

\subsection{Robustness and Reliability}
\vspace{-1.5mm}
Finally, the centralized framework is vulnerable to single-point failures. If the centralized model or the \ac{AI}/\ac{ML} Function fails, the entire \ac{MANO} procedures will be disrupted, leading to a significant degradation in network performance, or even malfunctions. To mitigate this challenge, the \ac{AI}/\ac{ML} Function must implement robust fault tolerance mechanisms to ensure the reliability of the centralized models and \ac{AI}/\ac{ML} Function. This may involve replicating the \ac{AI}/\ac{ML} Function across multiple \ac{Non-RT RIC} instances to ensure high availability and fault tolerance. Notably, this necessitates the establishment of external data/model importation mechanisms, as discussed in Scenario A, to enable warm/hot backups. By implementing these measures, the centralized framework can withstand single-point failures and maintain operational continuity, thereby safeguarding network performance and stability.

\section{Concluding Remarks and Future Outlook}\label{Sec:Conclusions}
\vspace{-1.5mm}
This article has presented three scenarios for integrating \ac{AI}/\ac{ML} algorithms into the \ac{SMO} framework. Through these scenarios, we have outlined how \ac{SMO} can evolve to address the burgeoning demands of managing the complex \ac{O-RAN} interfaces and components. Our focus then narrowed to a detailed exploration of one scenario, centralized \ac{ML}, where we have proposed an architectural solution. The proposal positioned the \ac{Non-RT RIC} as a central entity responsible for data collection as well as training, deploying, and refining the \ac{ML} model. However, the proposed solution presents a number of challenges that demand innovative solutions. Hence, we have identified several research challenges that require exploration to further enhance the intelligence of \ac{SMO}.

Regarding future work, we aim to delve into the exploration of Scenario A and Scenario C, as proposed in this article. In Scenario A, our focus lies on studying the mechanisms for data collection and model training from an external source. Once the model is prepared for deployment, the \ac{AI}/\ac{ML} Function may import it to incorporate the model into an optimization task. In Scenario B, our interest lies in exploring distributed learning techniques, particularly federated learning, tailored specifically for \ac{SMO}. The goal of this investigation could involve training and developing a global model for \ac{SMO}, wherein one management system, such as \ac{Non-RT RIC}, could act as a model aggregator, while other systems, such as \ac{3GPP-NSMS} and \ac{NFV-MANO}, serve as clients. By addressing these research challenges, future studies can contribute to the ongoing evolution of \ac{SMO}, driving innovation and efficiency in network and service \ac{MANO} within \ac{O-RAN}.
\vspace{-0.2mm}

\section*{Acknowledgment}
\vspace{-1.5mm}
This research was partially supported by the Horizon $\mathrm{2020}$ Research and Innovation Program of the European Union through Hexa-X-II under Grant $\mathrm{101015956}$.
\vspace{-0.2mm}

\bibliography{ref/mypaper01.bib} 
\bibliographystyle{ieeetr}
\end{document}